# Chapter 11 Students' interaction with and appreciation of automated informative tutoring feedback

Gerben van der Hoek[1], Bastiaan Heeren[2]†, Rogier Bos[3], Paul Drijvers[4], Johan Jeuring[5]

[1]https://orcid.org/0009-0004-0932-3065
[2]https://orcid.org/0000-0001-6647-6130
[3]https://orcid.org/0000-0003-2017-9792
[4]https://orcid.org/0000-0002-2724-4967
[5]https://orcid.org/0000-0001-5645-7681

**Abstract**

Computer aided formative assessment can be used to enhance a learning process, for instance by providing feedback. There are many design choices for delivering feedback, that lead to a feedback strategy. In an informative feedback strategy, students do not immediately receive information about the correct response, but are offered the opportunity to retry a task to apply feedback information. In this small-scale qualitative study, we explore an informative feedback strategy designed to offer a balance between room for exploration and mitigation of learning barriers. The research questions concern the ways in which students interact with the feedback strategy and their appreciation of error-specific feedback as opposed to worked-out solutions. To answer these questions, twenty-five 15-to-17-year-old senior general secondary education students worked for approximately 20 minutes on linear and exponential extrapolation tasks in an online environment. Data included screen captures of students working with the environment and post-intervention interviews. Results showed that room for exploration offered opportunities for self-guidance while mitigation of learning barriers prevented disengagement. Furthermore, students appreciated balanced feedback. We conclude that the balanced feedback strategy yielded fruitful student-environment interactions.

## 1 Introduction

Nowadays, computer-aided assessment in mathematics education is used more and more (Sangwin, 2015). For example, computer-aided assessment systems can aid the learning process by providing error-specific feedback. Designers of computer aided assessment systems face many choices with respect to feedback delivery. Narciss (2012) defines a feedback strategy as the specification of the way feedback is delivered within a learning environment. She singles out a specific feedback strategy, namely the informative tutoring feedback strategy (ITF strategy). In an ITF strategy, a student does not immediately receive information about the correct response, but is offered the opportunity to retry a task, to apply feedback information such as error specific hints. As such, an ITF strategy specifies a form of guidance for a student in the environment, where the student is provided opportunities to correct errors rather than study a correct solution.

There is an ongoing debate about the role of guidance during learning processes. In one corner we have, for instance, Kirschner et al. (2006) advocating direct instruction and worked examples. In the opposing corner, we have, for instance, de Jong et al. (2023) arguing in favour of inquiry-based approaches. Our goal is to use insights from this debate to set up an ITF strategy, which provides error-specific feedback and suitable subtasks to work on next. We aim to let students solve tasks without worked examples, but by constructing the solutions themselves. However, in some cases, students may have difficulty starting a task without direct instruction, therefore the environment offers optional direct instruction in the form of a video. In later stages, students can view worked-out solutions to the tasks.

In this chapter, we expand on our previous work about student experiences while working in an online environment (Van der Hoek et al., 2024) using new data. We designed an ITF strategy to support students in our learning environment. We analysed how this strategy affects the interactions of students with the environment, and how students appreciate the support they receive. Student behaviour while working in automated tutoring systems has been studied quantitatively, for instance, by Köck & Paramythis (2011) and Vaessen et al. (2014). Such quantitative studies rely on indicators to model students' behaviour, such as time spent in different modes of the environment or transition probabilities between these modes. However, such models of learners' interactions are an approximation of a complex reality. This is why, in this chapter, we use a qualitative approach to describe what happens when students are working in the environment.

A calculation by a student who is working in our environment is diagnosed only by inputting a final answer; this has two advantages. Firstly, working with final answer evaluation does not require additional skills to use the interface, allowing students to practice mathematics authentically (Kieran & Drijvers, 2006; Russell et al., 2003). And secondly, with final answer evaluation, there is no need for input fields to assess intermediate steps that could provide a scaffolding effect for the task (Tacoma et al., 2020). To provide error-specific diagnoses in our environment we use Model Backtracking (MBT) (Van der Hoek, 2022), a technique that diagnoses a final answer to identify errors a student has made throughout the computation. As such, it allows a student to work on a problem using pen and paper and input only their final answer, to receive feedback on the steps in their calculation.

To study how students interact with our design and whether students appreciate it, we use a qualitative small-scale design study with post-intervention interviews. We discuss how senior general secondary students aged 15 to 17 years interact with the environment. Furthermore, through the post-task interviews we investigate students' appreciation of different forms of feedback.

## 2 Theoretical framework

As with any learning, online learning requires guidance. In an online learning environment, this guidance can be provided as automated feedback. Below we discuss several feedback types relevant to an ITF strategy before moving to the role of guidance. We provide arguments from both sides of the discussion on guidance to argue that a balanced feedback strategy might be a good approach to facilitate learning in an online environment.

A feedback strategy specifies the delivery of feedback through a learning medium. In this study, we apply an ITF strategy (informative tutoring feedback strategy) (Narciss, 2012). An ITF strategy does not immediately present the correct response but offers learners the opportunity to retry tasks to apply previously received feedback information. For such feedback strategies, Narciss distinguishes three important dimensions, see Table 11.1: (1) the nature and quality of a feedback strategy, (2) the situational conditions of the instructional context, and (3) the individual characteristics of the learner. The first dimension, defining the nature of the strategy, includes three facets: (a) functional aspects related to objectives, such as fostering self-guidance and sustaining persistence; (b) aspects related to the purpose of the feedback content, such as identifying discrepancies between the learners' performance and the expected performance; (c) aspects related to the presentation of the feedback, such as the level of specificity. The second dimension, defining the instructional context, deals with the core features of the instructional approach. This includes, for instance, identifying learning obstacles and errors. In a broader scope, the instructional beliefs of the feedback designers are embedded in this dimension. The third dimension of learners' characteristics encompasses, for instance, prior knowledge and learning strategies. In what follows, we describe theory that will allow us to operationalise each of Narciss' three dimensions. In the design section, we will revisit these dimensions to elaborate how each is instantiated in our feedback strategy design.

| Narciss' dimensions of a feedback strategy |
|---|
| 1) Nature and quality of a feedback strategy<br>a) Functional aspects related to objectives<br>b) Aspects related to the purpose of the feedback content<br>c) Aspects related to the presentation of the feedback<br>2) Situational conditions of the instructional context<br>3) Individual characteristics of the learner |

Table 11.1 Dimensions of a feedback strategy according to Narciss (2012)

Shute (2008) and Narciss (2012) identify several types of feedback, see Table 11.2, such as verification, try-again, and elaborated feedback. Verification feedback provides learners with knowledge about a response's correctness, often referred to as knowledge of results (KR). Try-again feedback (TA) allows learners to provide a new response after some other type of feedback is provided. As for elaborated feedback, Shute distinguishes several variants, two of which are of interest here: Topic-contingent feedback and feedback on bugs. Topic-contingent feedback is feedback about the topic that is being studied, which could be a worked example (WE) of a task or direct instruction (DI); feedback on bugs is error-specific feedback (ES), which is based on a diagnosis of a learner's response. These five feedback types are incorporated into our environment.

| Abbreviation | Feedback type | Description |
|---|---|---|
| KR | Knowledge of results | Correct or incorrect responses are marked accordingly |
| TA | Try again | A student has the opportunity to try a task again after a failed attempt |
| WE | Worked example | A student can study a worked example to a task |
| ES | Error specific | A student receives feedback that is specific for an error made |
| DI | Direct instruction | The theory and procedures involved are explained by an expert. This could be a video recording. |

Table 11.2 Feedback types, abbreviations, and descriptions

When learners execute a task, they experience cognitive load on their working memory (Kirschner et al., 2006). This load can be reduced by using direct instruction or worked examples. For instance, Sweller et al. (1998) show that worked examples alleviated cognitive load for low-ability students. Studies on worked examples versus learning from solving problems in the previous century (Chi et al., 1989; Renkl, 1997) generally favour learning from worked examples. The cognitive load that task execution introduces can cloud the actual learning process. However, Chi et al. (1989) also reported that positive learning outcomes using worked examples strongly depend on a student's ability to self-explain the steps in the worked example. This shows that worked examples promote the learning process, but there is a danger in solely relying on them.

Learning is a form of self-development; it constitutes a positive change of behaviour and knowledge stored in the long-term memory; meanwhile, opportunities for self-development can be diminished by too much guidance. Moreover, through minimal guidance, students can develop the ability to evaluate their solution processes (Goodman & Wood, 2004). However, a drawback of minimal guidance is that a learner may experience a feeling of uncertainty (Bordia et al., 2004; Fedor, 1991) that could diminish their motivation. Uncertainty occurs when a learner feels they do not have sufficient information about their performance relative to task demands. This uncertainty can lead to frustration and disengagement (Williams, 1997). Thus, there should be enough room to explore in a learning environment, but there should also be possibilities to resolve uncertainty.

In conclusion: when learners perform a task, they can experience cognitive load and uncertainty, which can be reduced by using direct instruction and worked examples. However, too much guidance can impede self-development. To develop self-guidance abilities, there should be enough room for exploration for students. Cognitive demands and room for self-development must be balanced in an ITF strategy.

# 3 Research questions

Considering the theoretical framework, we set up the ITF strategy in our environment as follows: The available implemented feedback types, KR (Knowledge of Results), TA (Try-Again feedback), and ES (Error-Specific feedback), should allow students to postpone the use of the implemented WE (Worked Examples). As such, we create opportunities for students to develop self-guidance skills through exploration, and to receive guidance in case of cognitive overload or uncertainty. DI (Direct Instruction) is available in the form of a video should a student not know how to start.

We investigate whether our ITF strategy contributes to fruitful student-environment interactions and whether students appreciate it. We define an interaction as an event where the environment provides feedback and a student responds to the feedback by a subsequent action.

To do so, we address two research questions:

1. How do students interact with the various feedback types in the informative tutoring feedback strategy?
2. Do students appreciate the use of error-specific feedback as opposed to worked-out solutions?

# 4 Methods

We use a qualitative small-scale design study with post-task interviews. In this section, we describe the design, the instrument, the participants and treatment, as well as the data collection, and data analysis.

## 4.1 Design

The environment offers two topics: linear extrapolation and exponential extrapolation. These topics are part of the curriculum for Dutch senior general students in the social science stream. Students regularly make errors when solving tasks on these topics (Esteley et al., 2004; Van Dooren et al., 2005). Therefore, we designed an online environment for performing such tasks. The environment (in Dutch) is available through the following link: https://ideastest.science.uu.nl/mbt-server/.

Both the linear and exponential extrapolation topics have the same informative tutoring feedback strategy. We first elaborate on these common design features before elaborating on the topics' design. The common design is an operationalisation of Narciss' (2012) dimensions for feedback strategies; we will show how these dimensions are realised after elaborating on the common design.

When entering the environment, a student views an initial instructional video on how to work within the environment. Next, the student begins to work on the main task. After entering a result in the environment, it provides KR feedback, and when a more elaborate diagnosis of a student error is possible it provides ES feedback. A student then has the option to try again to obtain TA feedback. The ES feedback in the main tasks has low specificity (i.e., verbally formulated suggestions), whereas the ES feedback in the subtasks has higher specificity (i.e., suggestions that may contain calculations). WE feedback, a worked-out solution, is available only after a student has selected a subtask. A student, however, is at liberty to immediately select a subtask and view the worked-out solution. Once a student returns to the main task, WE feedback will be available for the main task. In the main task, a student can receive DI (direct instruction) using an instructional video on the topic. This video may be viewed only once to prevent a student from using it as a worked example. The flowchart in Figure 11.1 provides an overview of the various student options.

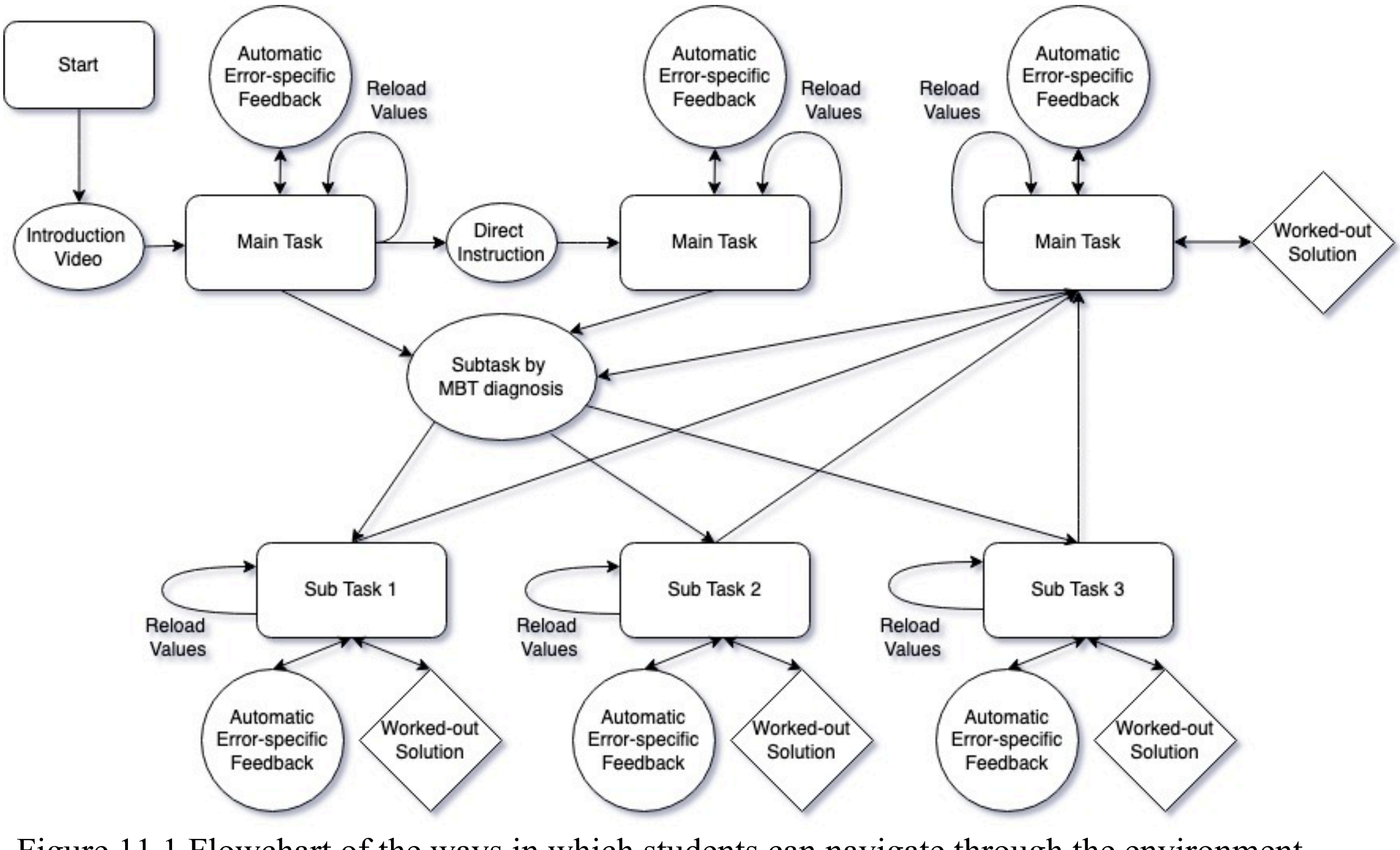

Figure 11.1 Flowchart of the ways in which students can navigate through the environment

Now we describe how the three dimensions of Narciss are implemented; later, this implementation will be made more concrete when we discuss the learning environment. We start by elaborating on the three aspects of the first dimension about the nature of the feedback strategy:

a) Our main functional objectives are fostering self-guidance and maintaining engagement. To achieve self-guidance, various feedback types besides worked-out solutions are offered to allow room for exploration. Cognitive load and uncertainty are reduced by subtasks (ST), direct instruction (DI) (i.e., an instructional video on extrapolation), or worked-out solutions (WE).

b) The feedback content aims to enable students to identify discrepancies between their performance and the expected performance. To achieve this, the feedback either provides information on the specific student input (i.e., KR, TA, and ES) or has a high level of specificity (i.e., DI and WE).

c) The feedback presentation is such that the specificity increases at a student's request: The ES feedback has higher specificity in the subtasks and a student has the option to view DI or WE.

Narciss' second dimension deals with the instructional context. In our case, we adopt opinions from both sides of the discussion between Kirschner et al. (2006) and De Jong et al. (2023) on guidance. This means that uncertainty and cognitive load are seen as the main learning obstacles and that worked examples should be avoided in the early stages to foster self-guidance. Moreover, allowing a student to choose a suitable form of help (i.e., ST, DI, WE), contributes to an ability to self-guide their learning process, although this is not further investigated in this study.

The third dimension encompasses learners' characteristics. In our case, the 15-to-17-year-old students have prior knowledge on the topics offered in the learning environment. However, it could be that this prior knowledge is insufficient to start a task, hence the environment offers direct instruction in a video. Furthermore, we hypothesise that these adolescents tend to quickly resolve uncertainty through worked examples. Accordingly, we remove the option to view a worked example at the start of the task and only allow to view the direct instruction video once. Below we describe the implementation of the two topics: linear and exponential extrapolation.

### *4.1.1 Linear extrapolation tasks*

For the linear extrapolation tasks, we set up a task format consisting of a main task and various subtasks, see Table 11.3. Students are first presented with the main task. When students input an erroneous solution for the main task, they receive feedback. After that, they can choose to correct their initial input or work on a subtask. In the latter case, the MBT system selects an appropriate subtask given the error.

The sets of numbers in each of the tables of the tasks below are so-called parameters. For each task, these parameters are randomly chosen from a set of 50 pre-calculated parameters that are tuned for final answer diagnosis accuracy using MBT techniques (Van der Hoek, 2022). This way a student can retry a task with different starting values, while final answer diagnosis accuracy is maintained.

<table>
<tr><th>Task</th><th>Example formulation</th><th>Additional Description</th><th>Learning goal</th><th>Subtask selected in case of…</th></tr>
<tr><td>Main linear extrapolation task</td><td>Given the table:<table><tr><td>$x$</td><td>23</td><td>85</td><td>97</td></tr><tr><td>$y$</td><td>15</td><td>41</td><td>?</td></tr></table>Use linear extrapolation to compute the value of the question mark</td><td>-</td><td>- Computing the slope (average rate of change)<br>- Using the slope to extrapolate</td><td></td></tr>
<tr><td>Subtask 1: Simpler numbers</td><td>Given the table:<table><tr><td>$x$</td><td>54</td><td>55</td><td>93</td></tr><tr><td>$y$</td><td>64</td><td>57</td><td>?</td></tr></table>Use linear extrapolation to compute the value of the question mark.<br>To do so, first compute the change of $y$ in a single step of $x$.</td><td>Task complexity is reduced compared to the main task since the $x$-coordinates of the given points are always consecutive and a way to start the computation is provided</td><td>- Computing the change of $y$ in this special case.<br>- Using the change of $y$ to extrapolate</td><td>- No input<br>- Undetectable error<br>- The student calculates:<br>$? = 41 + (41 - 15) = 67$</td></tr>
<tr><td>Subtask 2: Given slope</td><td>Given that the slope (rate of change) is equal to 8 for the following table:<table><tr><td>$x$</td><td>30</td><td>87</td></tr><tr><td>$y$</td><td>91</td><td>?</td></tr></table>Use linear extrapolation to compute the value of the question mark.</td><td>Task complexity is reduced relative to the main task by providing the slope</td><td>- Using the slope to extrapolate</td><td>- Correct calculation of the slope (rounding errors allowed) and detectable error elsewhere</td></tr>
<tr><td>Subtask 3: Computing slope</td><td>Given the table:<table><tr><td>$x$</td><td>35</td><td>62</td></tr><tr><td>$y$</td><td>47</td><td>68</td></tr></table>Compute the slope (the average rate of change)</td><td>Task complexity is reduced relative to the main task by requesting only the slope</td><td>- Computing the slope</td><td>- Detectable incorrect calculation of the slope (rounding errors allowed).</td></tr>
</table>

Table 11.3 Tasks design structure for the case of linear extrapolation

Figure 11.2 shows ES feedback a student receives when inversely computing the slope. To detect the various student errors, so-called buggy rules are implemented in the system. These buggy rules represent erroneous steps in a student's calculation (VanLehn & Brown, 1980).

The rules are based on work by Van Dooren et al. (2005) on the unwarranted use of proportional models in missing value problems. Subsets of the buggy rules for the main task were used for the subtasks.

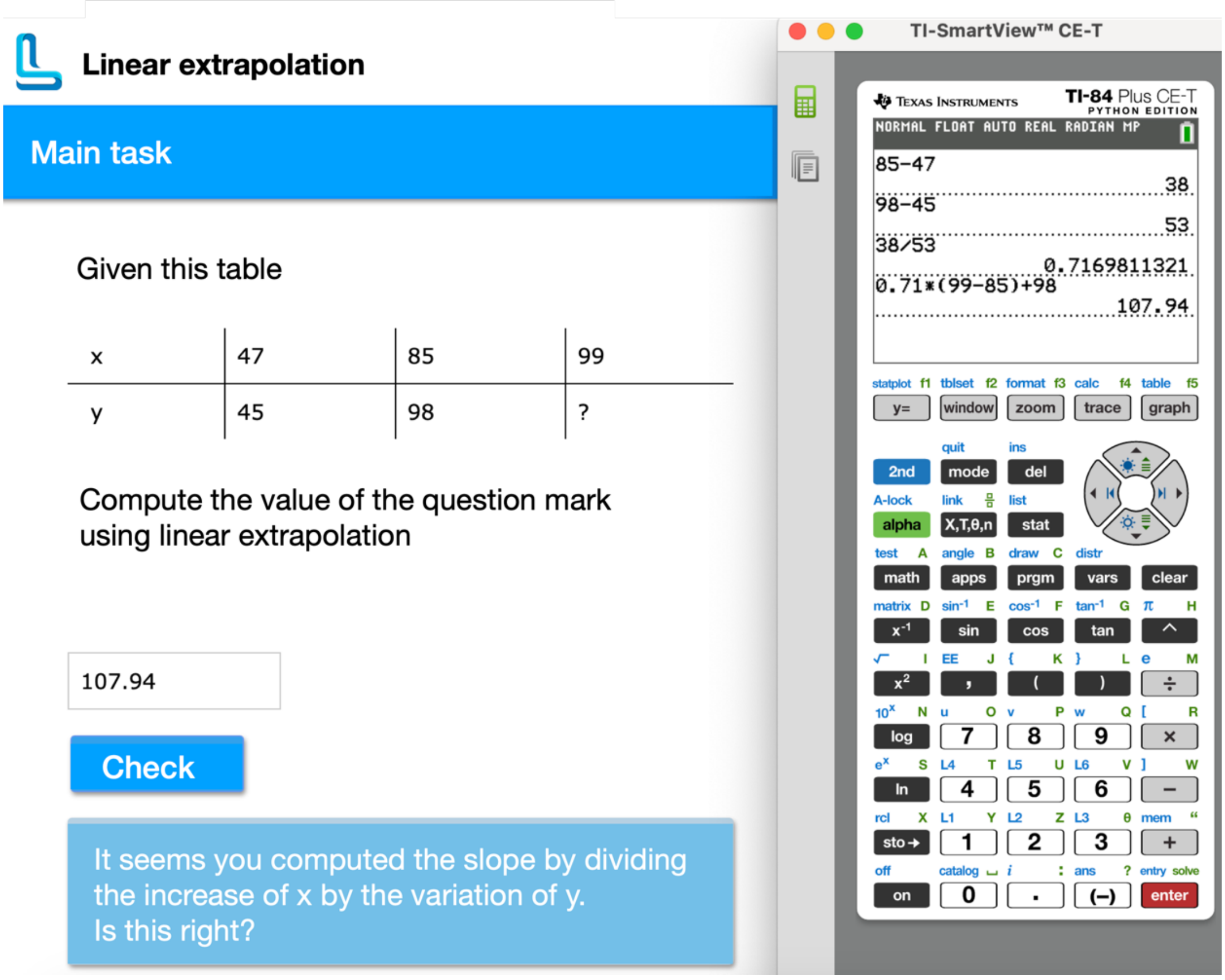

Figure 11.2 Example of error-specific feedback during a task

### *4.1.2 Exponential extrapolation tasks*

Analogous to linear extrapolation, we formulate a task with subtasks for exponential extrapolation in Table 11.4.

<table>
<tr><th>Task</th><th>Example formulation</th><th>Additional Description</th><th>Learning goal</th><th>Subtask selected in case of…</th></tr>
<tr><td>Main exponential extrapolation task</td><td>Given the table:<table><tr><td>$x$</td><td>77</td><td>80</td><td>85</td></tr><tr><td>$y$</td><td>58</td><td>55</td><td>?</td></tr></table>Use exponential extrapolation to compute the value of the question mark</td><td>-</td><td>- Computing the growth factor<br>- Using the growth factor to extrapolate</td><td>-</td></tr>
<tr><td>Subtask 1: Simpler numbers</td><td>Given the table:<table><tr><td>$x$</td><td>72</td><td>73</td><td>80</td></tr><tr><td>$y$</td><td>28</td><td>30</td><td>?</td></tr></table>Use exponential extrapolation to compute the value of the question mark.<br>To do so, first compute the growth factor for a single step of x.</td><td>Task complexity is reduced relative to the main task since the $x$-coordinates of the given points are always consecutive and a way to start the computation is provided</td><td>- Computing the growth factor in this special case.<br>- Using the growth factor to extrapolate</td><td>- No input<br>- Undetectable error<br>- The student calculates:<br>$? = \frac{55}{58} \cdot 55 = 52.155$</td></tr>
<tr><td>Subtask 2: Given growth factor</td><td>Given that the growth factor is equal to 1.059 for the following table:<table><tr><td>$x$</td><td>45</td><td>52</td></tr><tr><td>$y$</td><td>36</td><td>?</td></tr></table>Use exponential extrapolation to compute the value of the question mark.</td><td>Task complexity is reduced relative to the main task by providing the growth factor</td><td>- Using the growth factor to extrapolate</td><td>- Correct calculation of the growth factor (rounding errors allowed) and detectable error elsewhere</td></tr>
<tr><td>Subtask 3: Computing growth factor</td><td>Given the table:<table><tr><td>$x$</td><td>28</td><td>40</td></tr><tr><td>$y$</td><td>72</td><td>34</td></tr></table>Compute the growth factor for a single step of $x$.</td><td>Task complexity is reduced relative to the main task by requesting only the growth factor</td><td>- Computing the growth factor</td><td>- Detectable incorrect calculation of the growth factor (rounding errors allowed).</td></tr>
</table>

Table 11.4 Task design structure for the case of exponential extrapolation

For the exponential extrapolation task, the buggy rules are mainly based on work by Esteley et al. (2004) on using linear models in exponential situations. This environment was used to gather data on our research questions; in the next sections we explain our data analysis setup.

### 5.1 Instrument

Below, we present the post-task interview structure, see Table 11.5. Questions 1 through 7 are used to reflect on certain events during the session with the environment. They provide additional information for the first research question on the interaction with the environment. Questions 8 through 11 are used to gather information for future improvements of the environment. Finally, questions 12 and 13 are used to answer the second research question on the appreciation of an informative feedback strategy.

| Interview |
|---|
| 1. Did you understand the feedback you received? |
| 2. Do you understand your error now? |
| 3. Can you explain your error? |
| 4. Can you explain why it was an error? |
| 5. Was the subtask in line with the error you made? |
| 6. Did the subtask help you understand the main task? |
| 7. Did you understand the worked-out solution? |
| 8. What do you think about the environment? |
| 9. Are there cons to using this environment? |
| 10. Are there benefits to using the environment? |
| 11. What could be improved? |
| 12. What do you prefer, to receive feedback on a single error or to view a worked-out solution? |
| 13. Why? |

Table 11.5 Questions in the post-task interview

### 5.2 Participants and treatment

For this qualitative research, ten senior general secondary students from 10$^{th}$ grade and fifteen from 11$^{th}$ grade were recruited from six different classes in the school in the Netherlands where the first author is employed. Participation was based on availability and consent to partake.

The 10$^{th}$-grade students had received prior education on linear extrapolation as part of their standard curriculum. The 11$^{th}$-grade students had received prior education on linear and exponential extrapolation as part of their standard curriculum. However, the 11$^{th}$ and 10$^{th}$-grade students did not receive instruction on these topics in the four weeks before the experiment.

Students were invited to complete the main task in the environment in a single session. The 11$^{th}$-grade students could choose between linear and exponential extrapolation, or both when time allowed it. The 10$^{th}$-grade students worked on the linear extrapolation task only. Pen, paper, and an onscreen graphic calculator were available to the students. A researcher supported the students in case of confusion on how to operate the system, but not in case of confusion on the task. At times, the researcher reminded the students of the various options of the environment. After the end of the session with the environment, the researcher conducted a post-task interview with each student to determine the students' experiences with the environment.

### 5.3 Data collection

The data consist of screen capture recordings along with the voices of the students and the researcher. One recording was not saved properly, but the researcher provided a written account instead. Therefore, we have data from 25 students navigating the environment. The duration of the navigation sessions ranged from 10 to 25 minutes and was sometimes restricted due to external factors such as the start of the next class. Furthermore, we interviewed each student in 5 to 10 minutes sessions, which were audio recorded.

## 5.4 Data analyses

The data was analysed differently for the two research questions. For the research question on how the students interact with the ITF (informative tutoring feedback) strategy, units of analysis were identified. We define a unit of analysis or an event sequence as the student behaviour in between and including a starting state in Table 11.6 and a subsequent action in Table 11.7. The starting states and subsequent actions were chosen in this way because each navigating session can be covered by such units.

| Code | Current student state |
|---|---|
| KR | The student receives KR feedback but no ES feedback |
| ES | The student receives ES feedback |
| mES | The student receives ES feedback as a result of a misdiagnosis |
| WE | The student views the worked-out solution |
| ST | The student returns to the main task from a subtask |
| DI | The student views direct instruction |

Table 11.6 Codes that signify the start of a unit of analysis

| Code | Subsequent student action |
|---|---|
| IM | The student improves the input relative to the last input for the same task |
| nIM | The student does not improve the input relative to the last input for the same task |
| WE | The student views the worked-out solution |
| ST | The student chooses to work on a subtask |
| DI | The student views direct instruction |
| nCON | The student indicates not knowing how to continue |

Table 11.7 Codes that signify the end of a unit of analysis

If a student has no previous input, then any input is seen as an improvement. Additionally, improvement after viewing the worked example is measured through the first attempt with new starting values. Furthermore, when a student returns from a subtask, the last input in the main task is compared to the new input. A misdiagnosis occurs when the student input is not diagnosed properly.

If the allotted time expired during a unit of analysis, this unit was removed from the data. The transition frequencies from a student's state to a subsequent action are presented in Table 11.8. On 20% of the episodes resulting in either IM or nIM a second rater independently agreed fully with the initial coding. Furthermore, the second rater independently reviewed the excerpts coded with nCON and fully agreed.

To investigate students' self-guidance ability, the students' calculations were examined for signs of testing the correctness of an intermediate result. That is, a student uses values in the task formulation to check whether an intermediate calculation step is correct. The test itself, however, need not be correct. Excerpts of the sessions were coded for this event, see Table 11.9. A second rater reviewed these excerpts and fully agreed with the initial coding. From the episodes, six episodes were selected exemplifying certain typical or atypical behaviours.

For the research question on students' appreciation of error-specific feedback as opposed to worked-out solutions, we analysed the post-task interviews. These opinions were prompted by questions 12 and 13 in the post-task interview. First, we coded the utterances on error-specific feedback and worked-out solutions using data-driven coding. We then grouped these initial codes and merged them in an axial coding process until an overview of themes emerged, see Table 11.11. A second rater coded 20% of the excerpts; Cohen's kappa[1] was $\kappa = .69$. After this initial coding, both raters discussed the differences after which the raters agreed on this 20% of the data. Our sample size of 25 should be sufficient to achieve saturation of the themes in the interviews, because our population is fairly homogeneous and the research object (i.e., the response to questions 12 and 13) is narrowly defined (Hennink & Kaiser, 2022).

# 6 Results

In this section, we present the results of our study. We first present the results on the interaction with the feedback strategy, followed by the results on students' appreciation of error specific feedback opposed to worked examples.

## 6.1 Results on interactions

The results relevant to the first research question on the interactions in the environment are presented in Table 11.8.

| | | Student's next action | | | | | | |
|---|---|---|---|---|---|---|---|---|
| | | Improvement | Non-improvement | Worked Example | Sub task | Direct Instruction | Unable to continue | Total |
| Current feedback | Only knowledge of results feedback | 4 | 0 | 4 | 3 | 1 | 2 | 14 |
| | Error-specific feedback | 14 | 2 | 5 | 4 | 2 | 2 | 29 |
| | Misdiagnosis | 1 | 3 | 0 | 0 | 1 | 1 | 6 |
| | Worked example | 5 | 3 | 0 | 0 | 0 | 1 | 9 |
| | Subtask | 7 | 0 | 0 | 0 | 0 | 0 | 7 |
| | Direct Instruction | 13 | 0 | 0 | 1 | 0 | 0 | 14 |
| | Total | 44 | 8 | 9 | 8 | 4 | 6 | 79 |

Table 11.8 Frequencies of various event sequences

The improvement rate for direct instruction in this table is striking: 13 out of 14. This can be explained by the fact that direct instruction is often used at the start of a task, and no answer or a far-fetched answer is easy to improve upon. Another striking table entry is improvement through a subtask: 7 out of 7. This can be explained by the reduced cognitive load in the subtask in combination with the added presence of a worked-out solution allowing students to zoom in on their errors. Moreover, the subtask type is selected automatically based on the student error. The error-specific feedback shows improvement in roughly half the cases. In the other half of the cases, uncertainty is not removed sufficiently and students do not improve, seek additional feedback or even express an inability to continue.

[1] Multiple codes could be assigned to a single excerpt; therefore, an additional category (theme) was added to the codebook signifying a non-code. If the number of codes for an excerpt differed between raters, the smallest number of codes was supplemented with the non-code category, after which the number of codes for the excerpt is equal for both raters. Then Cohen's kappa was calculated by viewing each code assignment as a single observation, allowing for multiple observations using the same excerpt. Here, each time the raters assigned the same category to an excerpt, it was viewed as a single agreement; the order in which the categories were assigned was immaterial.

The results of the investigation on self-guidance are presented in Table 11.7. It shows that 1 in 5 students displayed self-guidance by checking intermediate results.

| Showed self-guidance | Number of students |
|---|---|
| No | 20 |
| Yes | 5 |
| Total | 25 |

Table 11.9 Self-guidance frequency

From the units of analysis in Table 11.6 we collected six episodes and explained them through known theory; an overview is provided in Table 11.8. These episodes were selected because they illustrate certain typical and atypical interactions, and because during the post-task interview students made comments shedding further light on what happened during the episode. The episodes show that, generally, the feedback strategy offers guidance specific to an individual student's needs; still, in some cases, the environment does not provide the required guidance.

| Episode | Interaction | Theoretical interpretation |
|---|---|---|
| *Ian, learning through a subtask* | Ian understood error-specific feedback on an error when he received it for the second time in the reduced setting of a subtask instead of in the main task. Upon returning to the main task, Ian solved it in a single attempt. In the post-task interview, Ian displayed insight into why his initial computation was erroneous. | Once the cognitive load (Sweller et al., 1998) was mitigated by the reduced complexity of the subtask, Ian could focus on his specific error. |
| *Dani, not knowing how to start* | Dani explicitly indicated she had no idea how to start; however, after receiving direct instruction she could complete a large portion of the main task. | Direct instruction (Kirschner et al., 2006) reduced cognitive load (Sweller et al., 1998) and produced immediate results. |
| *Leah, ES feedback* | Leah received error-specific feedback messages allowing her to complete the task without other forms of guidance. | Error-specific feedback (Shute, 2008) guided Leah towards a correct solution and produced direct results. |
| *Yuna, self-check* | Yuna tried to test whether an intermediate result was correct using values in the task, she explained in the post-task interview. As such, she exhibited self-guidance. | Exploration fostered a self-guiding ability (de Jong et al., 2023; Goodman & Wood, 2004) |
| *Eve, memorizing a worked example* | Eve unsuccessfully tried to memorize a worked-out solution to a task. She repeatedly returned to the worked-out solution and talked about memorizing it. However, she struggled to make sense of the task. | Eve lacked the proper self-explanation (Chi et al., 1989) skills needed to interpret a worked-out solution. |
| *Robert, uncertain* | Robert received knowledge of results feedback twice, but could not correct his errors; eventually, he sighed: "Now I have no idea what I did wrong." He ceased his activity until prompted by the researcher. | The feedback was insufficient to alleviate Robert's uncertainty (Bordia et al., 2004; Fedor, 1991) causing him to become disengaged (Williams, 1997), until prompted by the researcher. |

Table 11.10 Selected episodes and their theoretical interpretation

### 6.2 Results on student's appreciation of error-specific feedback

Concerning the results for the second research question on students' appreciation of error-specific feedback as opposed to worked examples, Table 11.11 shows the themes that resulted from the axial coding process. The themes *Combination* and *Self* have the highest frequency, showing that students appreciate a combination of first error-specific feedback and later worked-out solutions. This is in line with our ITF strategy setup. In the discussion section below, we will further analyse the contents of Table 11.11 in light of known theory.

| Themes | Description | Frequency |
|---|---|---|
| Combination | First specific feedback then later the worked solution | 13 |
| Self | If I view a worked-out solution I don't have to do anything myself OR If I don't view a worked-out solution I can still try it myself | 7 |
| Overview | A worked-out solution provides an overview OR Specific feedback does not provide an overview | 6 |
| How | Error-specific feedback does not always tell you how to proceed | 5 |
| Understand | After studying a worked-out solution, I feel I understand how to solve the task, but when I have to do the task again it turns out that I don't understand it | 5 |
| Pinpoint | In a worked-out solution, an error is not pinpointed OR Specific feedback pinpoints an error | 4 |
| Stop | After studying a worked-out solution, I don't want to continue, because I think I understand | 4 |
| Why | A worked-out solution does not always tell you why certain steps are made | 4 |
| Total | | 48 |

Table 11.11 Themes of students' opinions uttered during the post-task interview

## 7 Discussion

With respect to the first research question on the interaction with the informative feedback strategy (Narciss, 2012), we conclude from the results that the environment functions as desired. In Table 11.8 the total number of times a student either did not improve or indicated an inability to continue is 14. This leaves a total of 65 out of 79 successful transitions or improvements, where cognitive load (Sweller et al., 1998) and uncertainty (Bordia et al., 2004; Fedor, 1991) were such that students could seek additional guidance or improve their results. Moreover, Table 11.9 shows that, on average, one in five students showed self-guidance abilities.

In Table 11.10, Ian, Dani, Leah, and Yuna show how the informative feedback strategy guides the learning process. Furthermore, Table 11.8 and Table 11.9 show that these episodes can be seen as fairly typical since improvement occurred 44 out of 79 times. The episodes of Eve and Robert show that despite the efforts to remedy the lack of self-explanation skills (Chi et al., 1989) and prevent disengagement due to uncertainty (Bordia et al., 2004; Fedor, 1991; Williams, 1997), these effects still occur in the interaction with the environment. However, Table 11.8 shows that these events are somewhat atypical, since non-improvement occurred only 8 out of 79 times and inability to continue 6 out of 79. Perhaps adding a prompt suggesting the use of direct instruction after several incorrect inputs could remedy this issue.

As for the second research question on students' experiences with error-specific feedback and worked-out solutions, the results uncover two possible dangers of starting early on with worked-out solutions. Firstly, based on the *Why* theme, with early use of worked-out solutions students might not be ready to self-explain them (Chi et al., 1989). Secondly, from cases coded as *Stop* and *Understand*, we observe students can gain a false sense of certainty. This unjustified certainty causes them to be satisfied with the state of affairs and cease their learning activities, only to find out later they do not yet meet the requirements to solve similar tasks. This provides another argument to be careful with worked-out solutions during the learning process (de Jong et al., 2023).

The themes most mentioned are *Combination* and *Self*. They show that students appreciate a combination of first error-specific feedback and later worked-out solutions. This aligns with the main design principle of our feedback strategy where room for exploration allows students to develop self-guidance abilities. This indicates that students appreciate a balanced informative feedback strategy.

There are some limitations to the validity and generalisability of this study. The sample size is small and the participants came from the same school in the Netherlands. This does not contribute to the generalisability of the findings. Furthermore, we do not measure any learning effects; but rather, analyse students' appreciation of the learning strategy. We do however offer explanations from known theory for the phenomena, which somewhat contributes to generalisability. Cultural influence and the inquiry-based book series used at the school may have influenced the results with respect to the second research question.

# 8 Conclusion

In conclusion, the results show that the ITF (informative tutoring feedback) strategy (Narciss, 2012) that balances guidance to accommodate students' individual needs is a strategy that students appreciate, and shows fruitful interactions. However, it remains a question of how this balanced ITF strategy would perform relative to a feedback strategy that, for instance, only provides worked-out examples and knowledge of results. Future research can shed light on such questions. Nonetheless, in this study, we showed that allowing room to explore (Goodman & Wood, 2004) while keeping cognitive load and uncertainty (Bordia et al., 2004; Fedor, 1991; Sweller et al., 1998; Williams, 1997) manageable promotes the learning process. The exploration fosters a self-guidance ability, while managing cognitive load and uncertainty allows students to improve themselves and remain engaged.

This shows that, for 15- to 17-year-old students, automated formative assessment benefits from a balanced informative tutoring feedback strategy for two reasons. Firstly, it allows room for exploration, but can still provide the guidance needed for the individual student. Secondly, students appreciate receiving feedback through such a strategy. For developers of computer-aided formative assessment systems for this age group, these are important results to consider when deciding on an appropriate feedback strategy.

# Acknowledgement

We are very sorry that co-author Bastiaan Heeren passed away before the publication of this chapter, and are grateful for his ability to articulate our thoughts, sometimes even before we had them.